# Relaxation characteristics and spectral compression in the system of coupled radiating oscillators

**Henrik Parsamyan,* Hasmik Shahinyan, and Khachatur Nerkararyan**

*Institute of Physics, Yerevan State University, A. Manoogian 1, Yerevan, 0025, Armenia*

Corresponding author*:* *hparsamyan@ysu.am*

The electromagnetic response of two identical coupled oscillators to an ultrashort pulse is investigated. Due to interference between the radiation emitted by the constituent oscillators, the temporal dynamics of the dipole radiation from the system are governed by a nontrivial relationship between their dipole moment amplitudes and differ markedly from those of an isolated oscillator. An analysis based on coupled-mode theory demonstrates that, under appropriate conditions, the energy of an incident ultrashort pulse can be converted into quasi-monochromatic radiation with high efficiency. The theoretical results are supported by qualitative numerical simulations of thin conductive rods resonant in the terahertz frequency range and separated by a deeply subwavelength distance, providing a physical realization of the coupled radiating-oscillator system.

## Introduction

The resonant interplay of electromagnetic waves with the structured matter lies at the heart of many wave-matter interaction phenomena, governing the spatiotemporal characteristics of scattered, transmitted or reflected waves and the formation of tailored spectral responses[1–4]. Advances in fabrication techniques over the past two decades have significantly accelerated progress in modern photonics and metasurface research, enabling the integration of multiple subwavelength metallic and/or dielectric elements featuring resonant responses into a metasurface design. Electromagnetic coupling between the resonant elements – either within the same unit cell or between adjacent unit-cells in bilayer configurations – has unveiled new spectral features[5].

This shift from isolated to coupled resonant elements is not merely quantitative but also qualitative. While the interaction of electromagnetic waves with an isolated subwavelength element is well-characterized by standard scattering theories[6,7], architectures incorporating multiple resonant objects facilitate access to significantly higher levels of spectral and hence – modal sophistication arising from their mutual electromagnetic coupling[8–11]. In complex systems, the response evinces as an emergent property of the coupled ensemble, exhibiting spectral characteristics far more intricate than the fundamental resonances of a single constituent[12–14]. Electrodynamic near-field coupling and interference between resonant regimes have enabled several prominent phenomena, most notably Fano interference[15–17] and classical analogues of electromagnetically induced transparency (EIT)[2,18]. The coupled systems composed of two resonant units support hybridized modes with different radiation characteristics. The symmetric mode, also referred to as the antibonding or bright mode, is characterized by a strong net dipole moment arising from the in-phase oscillations of the resonators. In contrast, the antisymmetric mode, also referred to as the bonding or dark mode, has a zero net dipole moment due to the out-of-phase oscillations of the resonators[19–21]. These distinct dipole characteristics directly determine the radiative properties of the two modes: the symmetric mode exhibits a broad linewidth and, consequently, manifests as short-lived oscillations, whereas the antisymmetric mode is characterized by a narrow linewidth and manifests as long-lived oscillations, reflecting their disparate radiative decay rates and relaxation dynamics[22]. The emergence of these hybridized modes introduces additional degrees of freedom, enabling engineering of the spectral and spatial optical response. In the framework of coupled-mode theory[23,24], the global spectral features are governed primarily by the coupling coefficient. While the relaxation dynamics of an isolated resonant element are dominated by intrinsic material absorption and radiative damping, the introduction of a long-lived "antisymmetric" mode reveals a nontrivial functional dependence between the total decay rate of the system and the inter-element coupling. Nevertheless, the fundamental physical aspects by which the coupling strength governs the relaxation dynamics of these hybridized modes, thereby determining their spectral characteristics, often remain overlooked.

Importantly, in addition to the structural geometry and intrinsic material parameters of resonant elements, the characteristics of the excitation pulse constitute a decisive factor in engineering its optical response, as they directly govern the temporal dynamics of charge carriers and molecular polarization in matter[25]. In this context, ultrafast optics, employing few-cycle ultrashort pulses for probing photonic structures, provides a powerful framework for tailoring and controlling the resulting spectral response. The interplay between ultrashort pulse excitation and inter-element coupling within a unit-cell array can give rise to nontrivial dynamical effects, one of which is the central focus of the present work.

Recently, the interaction of a system of two coupled dipole-active oscillators with an ultrashort THz pulse was investigated in[26]. The energy of the resulting converted monochromatic wave is found to noticeably exceed the energy within the corresponding spectral region carried by the incident pulse. It is evident that this effect cannot be explained by conventional wave filtering and requires a detailed analysis, necessitating an approach that goes beyond the linear theory of coupled modes.

In this paper, we theoretically investigate the electromagnetic response and mode formation peculiarities of a system consisting of two identical coupled oscillators in response to an ultrashort pulse. Owing to the dipole-active nature of the individual oscillators, radiative decay constitutes the dominant loss channel in the system. The subwavelength separation between the oscillators enables strong electromagnetic coupling, giving rise to characteristic coupled modes. In-phase oscillations of the oscillators lead to the formation of a symmetric mode with a broad radiation linewidth, whereas out-of-phase oscillations give rise to an antisymmetric mode with a strongly reduced net dipole moment and, consequently, a much narrower radiation linewidth [14,27,28]. As a result, the relaxation parameters associated with the symmetric and antisymmetric modes differ significantly. This distinction is important not only for describing the relaxation dynamics of the resulting modes but also because these parameters govern the mode formation process itself. In particular, when only one oscillator in the coupled system is excited by an external stimulus, it remains unclear which relaxation parameter can universally describe the dynamics of mode formation. This work aims to clarify this issue and investigate the nontrivial phenomena arising from the coupling between the oscillators.

**Theoretical model**

Let the ultrashort Gaussian pulse, impinging on the system of the two identical coupled rods and interacting with only one of the rods, be represented by the following expression:

$$s_i(t) = s_{i0} \exp\left[-\left(\frac{t-t_0}{\tau}\right)^2\right] cos\omega t, \tag{1}$$

where $\omega$ is the carrier angular frequency, $\tau$ is the pulse length, $t_0$ is the time delay, and $|s_{i0}|^2$ represents the power carried by the incident wave [23]. In the following analysis, we will omit the complex quantities in the equations for convenience, writing the $P_1(t)$ and $P_2(t)$ electric-dipole moments of each oscillator as follows:

$$P_1(t) = p_{11}(t) cos\omega t + p_{12}(t) sin\omega t, \tag{2}$$

$$P_2(t) = p_{21}(t) cos\omega t + p_{22}(t) sin\omega t. \tag{3}$$

Here the dipole moment amplitudes $p_{ik}$ $(i = 1; 2,\ k = 1; 2)$ are normalized such that $(p_{ik})^2$ equals the wave energy accumulated by each individual rod with the corresponding oscillator law (cos$\omega t$ or sin$\omega t$). Therefore, the energy of the oscillation corresponding to cos$\omega t$ can be determined as $(p_{11}^2 + p_{21}^2)$, and the dipole moment – as $(p_{11} + p_{21})$. Similarly, for oscillations of the system following the sin$\omega t$ law, the energy is determined by the expression $(p_{12}^2 + p_{22}^2)$, while the total dipole moment is given by $(p_{12} + p_{22})$. Notably, in the considered system of coupled rods the separation between identical oscillators is deeply subwavelength, whereas their coupling energy is much smaller than the wave energy accumulated in an individual oscillator.

Thus, starting from the equations of the coupled oscillators and transitioning to the slowly varying amplitude approximation, within the framework of temporal coupled-mode theory, we obtain the following system of equations:

$$\dot{p}_{11} + (\gamma_{dc} + \gamma_e)p_{11} + \Omega p_{12} + \mu p_{22} = 0, \tag{4}$$

$$\dot{p}_{12} + (\gamma_{ds} + \gamma_e)p_{12} - \Omega p_{11} - \mu p_{21} = \sqrt{2\gamma_0} s_{i0} \exp\left[-\left(\frac{t-t_0}{\tau}\right)^2\right], \tag{5}$$

$$\dot{p}_{21} + (\gamma_{dc} + \gamma_e)p_{21} + \Omega p_{22} + \mu p_{12} = 0, \tag{6}$$

$$\dot{p}_{22} + (\gamma_{ds} + \gamma_e)p_{22} - \Omega p_{21} - \mu p_{11} = 0, \tag{7}$$

with

$$\gamma_0;\ \gamma_e;\ |\Omega|;\ \mu \ll \omega_0. \tag{8}$$

Here $\gamma_0$ is the damping frequency of the individual isolated oscillator due to the dipole radiation, $\gamma_e$ is the damping coefficient associated with the non-dipolar radiation losses, $\Omega = \omega - \omega_0$, $\omega_0$ is the resonance frequency of the single oscillator. The parameter $\mu$ is the coupling coefficient, while the radiative damping rate of the system was partitioned into $\gamma_{ds}$ and $\gamma_{dc}$, corresponding to the *sine* and *cosine* components of the dipole-moments, respectively. In the presented system of equations, the variables $p_{ik}$ are assumed to change significantly over time, and the strong-coupling approximation is realized. The full derivation of the governing differential equations within the framework of coupled-mode theory, together with the theoretical derivation and identification of the symmetric and antisymmetric modes, is presented in Supplementary Information, Section S1.

Of fundamental importance in this system is the determination of the relaxation coefficient associated with dipolar radiation. Generally, it is defined as the ratio of the radiated power to the total energy content of the oscillator[23]. Given that the power of dipole radiation scales with the square of the dipole moment, under the current considerations the relaxation coefficients exhibit a pronounced non-linear form and for the cosine and sine oscillations are determined by the following expressions:

$$\gamma_{dc} = \gamma_0 \frac{(p_{11} + p_{21})^2}{(p_{11}^2 + p_{21}^2)}, \qquad \gamma_{ds} = \gamma_0 \frac{(p_{12} + p_{22})^2}{(p_{12}^2 + p_{22}^2)}, \tag{9}$$

The derivation of the relaxation coefficient is provided in detail in Supplementary Information, Section S2.

The process under investigation occurs within a relatively narrow frequency range near $\omega_0$, allowing $\omega_0$ to be used as the normalization frequency in the coefficient for $p_{ik}$ in all cases. When numerically solving the system of equations (4) – (7), we use the dimensionless time $t' = \gamma_0 t$. The theoretical results were obtained using the following parameters: $\mu = \gamma_0$, $\Omega = \gamma_0$, $\gamma_e = 0.1\gamma_0$, $\tau = 0.9/\gamma_0$ and $\omega_0 = 5\pi\gamma_0$.

**Results and discussion**

To establish a direct correspondence between the theoretical model presented above and a realistic electromagnetic system, we consider an equivalent configuration consisting of two parallel conductive rods separated by a deeply subwavelength distance that is also smaller than the rods' radius, thereby enabling strong electromagnetic coupling and the formation of coupled modes. One of these modes is the dipole-active symmetric mode, characterized by a broad radiation linewidth, whereas the other is the dipole-passive antisymmetric mode, characterized by a considerably narrower radiation linewidth [20,29]. This difference arises because the surface charges and their associated electromagnetic fields in the adjacent rods oscillate in-phase for the symmetric mode and out-of-phase for the antisymmetric mode. Consequently, when only one of the coupled rods is selectively excited, its dipole-active nature enables efficient energy accumulation, followed by redistribution of the accumulated energy between the coupled modes. The interaction of the coupled rods with an ultrashort pulse is numerically investigated using the Lumerical FDTD environment. Further details on the numerical simulations and model setup are provided in Section 3 of the Supplementary Information. The geometry of the structure and the temporal profile of the incident field are shown in Fig. 1(a).

The choice of geometry of oscillators and spectral parameters is justified as follows: (*i*) a thin conductive rod of finite length ($D \ll L$) acts as an open resonator supporting Sommerfeld surface standing waves [30], (*ii*) Ohmic losses in metals are negligible in the THz spectral range, and (*iii*) in terms of the ratio between the wave period and pulse duration, terahertz ultrashort pulses approach the practical limit. This configuration serves as a physical realization of the coupled oscillator model introduced above, where the strong coupling between the rods is achieved by placing them at a deeply subwavelength separation, of the order of the rod radius. In what follows, theoretical predictions and numerical results are presented in parallel in order to validate the analytical framework and elucidate the underlying dynamical mechanisms. Unless stated otherwise, the theoretical results were obtained using the following parameter values: $\mu = \gamma_0$, $\Omega = \gamma_0$, $\gamma_e = 0.1\gamma_0$, $\tau = 0.9/\gamma_0$ and $\omega_0 = 5\pi\gamma_0$. The condition $\mu = \Omega$ corresponds to the excitation at the resonance frequency of the antisymmetric mode. Within the theoretical description, the presence of nontrivial relaxation coefficients in equations (4)-(7) gives rise to two qualitatively different dynamical regimes of the oscillation amplitudes in time (Fig. 1(b) and (c)). During the initial time interval, with a characteristic duration on the order of $1/\gamma_0$, the oscillation amplitudes rapidly decay toward a distinct nonzero value. This process, governed by losses due to dipole radiation, results in self-organization of the system. This regime is followed by the second time period, starting when the total dipole moment vanishes and $\mathrm{p}_{11}(t) = -\mathrm{p}_{21}(t)$, $\mathrm{p}_{22}(t) = -\mathrm{p}_{12}(t)$. In this time period, the radiation losses are strongly suppressed, and the system is characterized by slowly decaying oscillations, approximately scaling as $exp(-\gamma_e t)$. In other words, only the antisymmetric mode of the coupled-oscillator system persists, featuring charges and their accompanying electromagnetic fields oscillating out-of-phase. The time evolution of the oscillation amplitudes in the numerical model was determined under conditions where the losses due to dipole radiation significantly exceed the non-dipolar losses ($\gamma_0 \gg \gamma_e$). Two point monitors [Points A and B in Fig. 1(a)] were placed at carefully selected positions near the outer top edges of the rods, away from the gap region, to track the temporal evolution of the individual rod contributions to the hybrid modes. The corresponding results for the $E_x$-components are presented in Fig. 1(d) and (e). Both waveforms exhibit an initial short-lived transient followed by a long-lived oscillatory regime, in close agreement with the theoretical dipole amplitudes $P_1$ and $P_2$ of the individual oscillators in the coupled system. We now evaluate the temporal dynamics of the waveforms associated with the hybridized modes of the coupled oscillators using both analytical and numerical approaches.

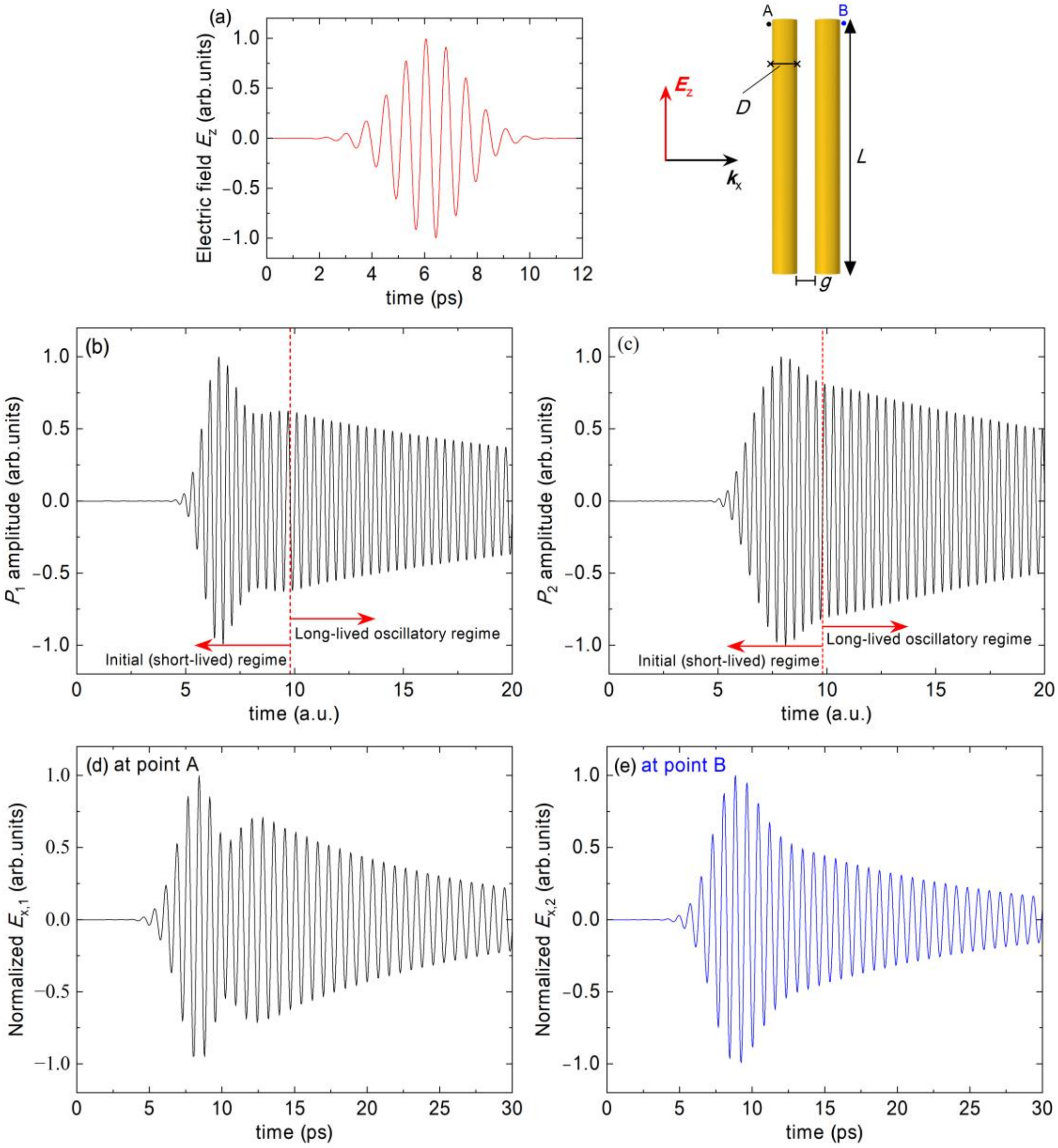

Fig. 1. (a) The incident pulse and the schematic of the numerical model of coupled oscillators, represented as thin conductive rods. $D$ and $L$ denote the diameter and length of the rods, respectively, and $g$ is the surface-to-surface separation between them. Points A and B mark the positions of time monitors used to evaluate the temporal dynamics of each coupled rod. (b-c) Temporal dynamics of normalized theoretical electric dipole moments (b) $P_1$ and (c) $P_2$ of the two oscillators in the coupled system, calculated using equations (2) and (3). The red line marks the approximate boundary between initial (short) and long-time oscillatory regimes. (d-e) Temporal dynamics of the normalized $x$-component of the electric field ($E_{x,i}$, $i$ = 1; 2) of the (d) first and (e) second rods in the coupled system. Two point-like monitors were introduced to separately probe the local electric field of each oscillator in the coupled system. The monitors were positioned near the outer edge of the top facet of each rod, outside the inter-rod gap region. Diameter and length of rods, and the distance between them are 10 μm, 100 μm and 3 μm , respectively. Theoretical results were obtained for the following parameters: $\mu = \gamma_0$, $\Omega = \gamma_0$, $\gamma_0 = 1$, $\gamma_e = 0.1\gamma_0$, $\tau = 0.9/\gamma_0$ and $\omega_0 = 5\pi\gamma_0$.

In the former, the wave fields associated with the symmetric and antisymmetric modes are defined as $P_s = P_1 + P_2$ and $P_{as} = P_1 - P_2$, respectively, where $P_1$ and $P_2$ are obtained from Eqs. (2) and (3) by solving the system of equations (4)-(7). Note that the values of $P_1$ and $P_2$ can always be represented as a superposition of symmetric $P_s$ and antisymmetric $P_{as}$ modes of the structure.The normalized theoretical temporal evolutions of the dipole moments associated with the symmetric ($P_s$) and antisymmetric ($P_{as}$) modes are presented in Fig. 2(a) and (b), respectively. The symmetric mode exhibits a comparatively short lifetime, and its temporal envelope closely follows that of the incident pulse. This behavior indicates strong radiative damping, as once the external driving field vanishes, the stored energy is rapidly released into the far field. In contrast, the antisymmetric mode demonstrates qualitatively different dynamics. During the initial stage, the oscillation amplitude increases rapidly under the action of the ultrashort excitation pulse. After the driving field has passed, however, the coupled oscillators sustain long-lived oscillations, indicating a substantially reduced radiative decay rate. These trends are corroborated by full-wave FDTD simulations performed with the parameters specified in the discussion of Fig. 1.

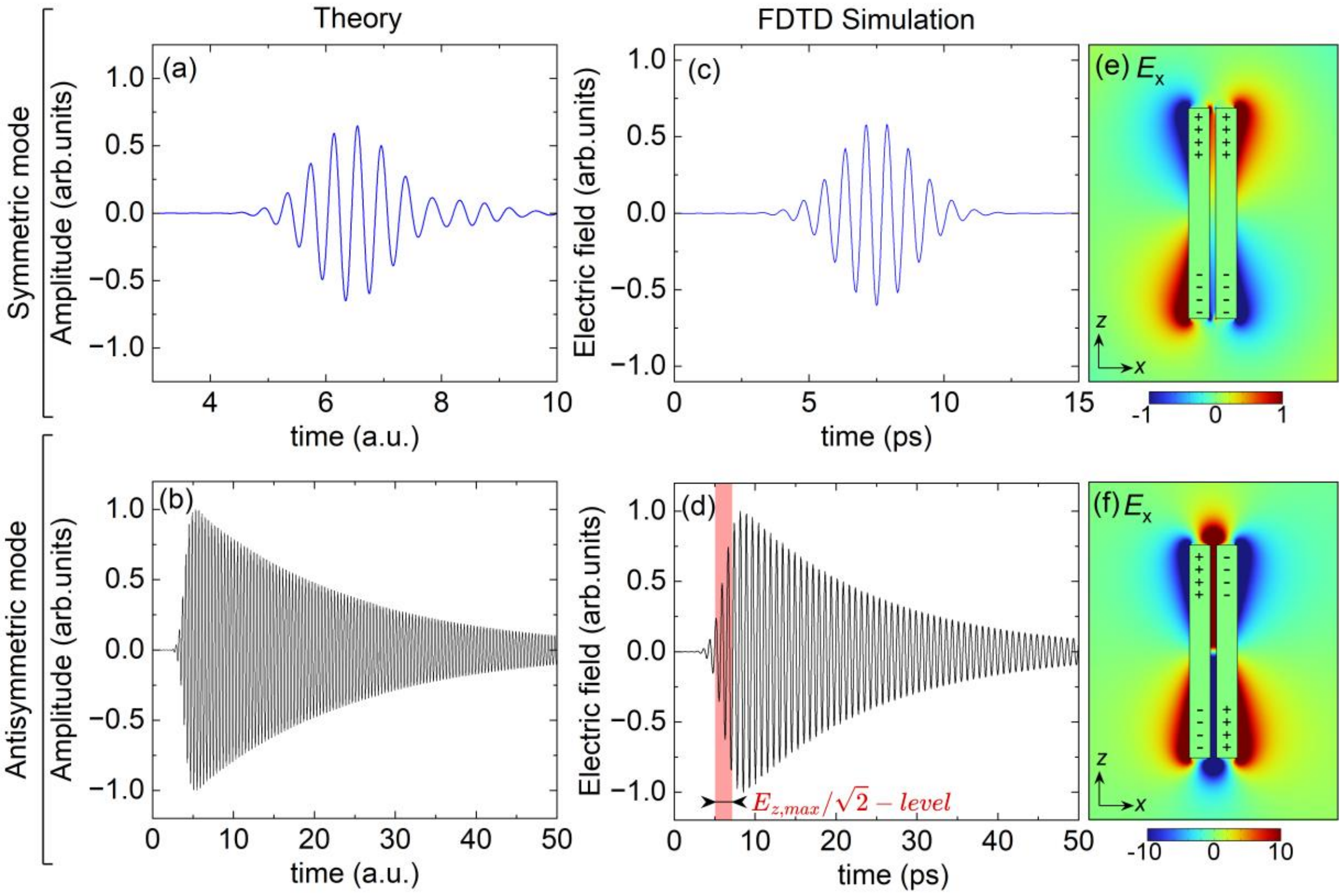

Fig. 2. (a, b) Theoretically calculated temporal evolution of the dipole moment amplitude for the (a) symmetric and (b) antisymmetric modes. (c, d) Corresponding FDTD-simulated electric field waveforms for the (c) symmetric and (d) antisymmetric modes. Theoretical results were obtained for the following parameters: $\mu = \gamma_0$, $\Omega = \gamma_0$, $\gamma_e = 0.1\gamma_0$, $\tau = 0.9/\gamma_0$, and $\omega_0 = 5\pi\gamma_0$. Red shaded area in (d) shows the $E_{z,max}/\sqrt{2}$-level of the incident pulse. (e, f) Distributions of the electric field $E_x$ component in the $xz$-plane and the corresponding charge distributions of the (e) symmetric and (f) antisymmetric modes.

To excite the symmetric mode in the numerical model, the excitation configuration was modified so that the incident pulse, while maintaining its original polarization ($E_z$), simultaneously illuminates both rods along the y direction ($k_y$). The FWHM of $|E_x|^2$ for the symmetric mode is approximately 2 ps, confirming pronounced radiative losses [Fig. 2(c)]. The temporal evolution of the $E_x$-field corresponding to the antisymmetric mode is shown in Fig. 2(d). The numerical results are in good agreement with the theoretical predictions and reveal long-lived oscillations. The low relaxation rate of this mode originates from the strongly suppressed net dipole moment caused by the out-of-phase oscillations of charge and their associated electric fields. This leads to strong near-field coupling and pronounced electromagnetic energy localization in the inter-rod region. The distributions of the normalized $E_x$ component of the electric field for the symmetric and antisymmetric modes in the lateral $xz$-plane of the system, together with the corresponding surface charge distributions on the rods, are shown in Fig. 2(e) and (f), respectively. The field distribution of the symmetric mode indicates its nearly pure dipolar nature, with the rods oscillating in-phase. In contrast, the field distribution of the antisymmetric mode reveals out-of-phase oscillations of the rods and a quadrupolar charge distribution [26]. This results in strong electromagnetic coupling between them, which is manifested by a pronounced field enhancement in the interrod region.

The normalized power spectra $P_\omega$ of the symmetric and antisymmetric modes in the coupled rods system are plotted in Fig. 3(a) as blue and black curves, respectively, and the green line shows that of the single rod. The spectra of symmetric and single rod modes are scaled by a factor of 100 to facilitate comparison of the two curves. It is seen that the FWHM of the symmetric and single rod modes are comparable, being approximately 0.15 THz and 0.18 THz, respectively. These linewidth values are also comparable with the bandwidth of the incident pulse. In contrast, the antisymmetric mode exhibits a significantly narrower linewidth of 0.022 THz. Such a sharp reduction in the relaxation rate compared with the single-rod response arises from the nontrivial dependence of the radiative damping on the dipole amplitudes under interference conditions.

The system of equations (4)-(7) also enables the evaluation of the wave energy accumulation, redistribution and persistence in the coupled oscillator system. The accumulated energy is defined as:

$$W_{tot} = \frac{\sqrt{2}}{2}(p_{11}^2 + p_{12}^2 + p_{21}^2 + p_{22}^2), \tag{10}$$

the normalized temporal evolution of which at frequencies of $\Omega = 0.5\gamma_0$, $\gamma_0$ and $1.5\gamma_0$ are plotted in Fig. 3(b), from bottom to top. The red circles show the exponential fitting of the long-lived portion of the oscillations using the function $W_0\times\exp(-2\gamma_e t)$. The fitting yielded $\gamma_e = 0.08\gamma_0$ and $W_0 = 0.7$ for $\Omega = 0.5$, $\gamma_e = 0.049\gamma_0$ and $W_0 = 1.55$ for $\Omega = \gamma_0$, and $\gamma_e = 0.062\gamma_0$ and $W_0 = 1.45$ for $\Omega = 1.5\gamma_0$. This behavior is further illustrated in Fig. 3(c), where the normalized total energy accumulated in the system is presented as a function of time and frequency.

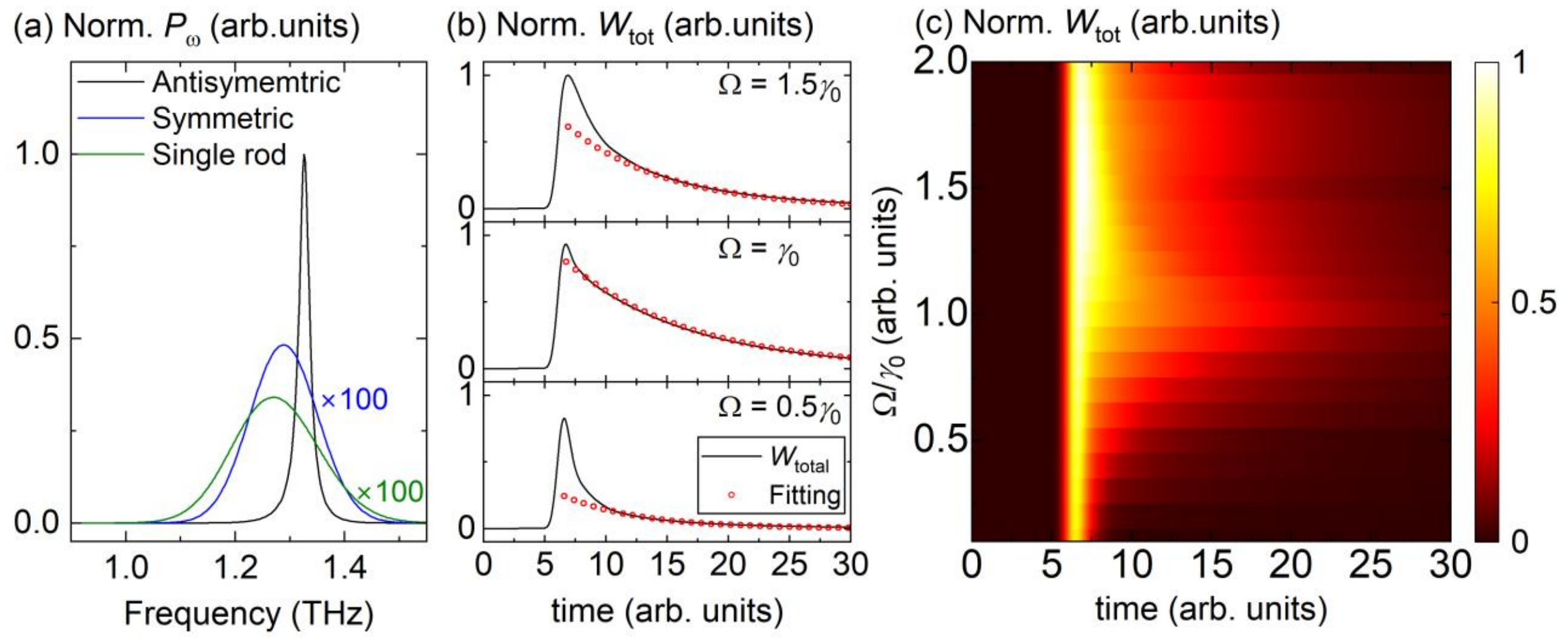

Fig. 3. (a) Normalized power spectrum of the symmetric (blue) and antisymmetric (black) modes in the coupled rods; and single rod (green). The symmetric and single rod mode spectra are scaled by a factor of 100 for comparison. (b) Temporal evolution of the total energy of the coupled oscillators for the frequency of the incidence field of $\Omega = 0.5\gamma_0$, $\gamma_0$ and $1.5\gamma_0$. The red circles indicate the exponential fitting of the long-lived part of the oscillations. (c) Color map of the theoretically calculated total energy as a function of normalized time and resonance frequency $\Omega/\gamma_0$. The remaining parameters are fixed at $\mu = \gamma_0$, $\gamma_e = 0.1\gamma_0$, $\tau = 0.9/\gamma_0$, and $\omega_0 = 5\pi\gamma_0$.

The energy map clearly reveals two temporal intervals characterized by distinct relaxation dynamics. In the coupled-rod system under the chosen excitation configuration, the incident pulse first couples to the front rod. The relatively broad linewidth of the single-rod response enables effective energy transfer into the system. The injected energy subsequently redistributes between the hybridized symmetric and antisymmetric eigenmodes of the coupled structure. The initial interval corresponds to the energy build-up driven by the incident ultrashort pulse. During this stage, the system absorbs and redistributes energy between the coupled modes of the oscillators under direct external ultrashort pulse excitation. The subsequent interval represents the free evolution of the system after the excitation pulse has passed. In this regime, the stored energy is converted into long-lived, highly monochromatic oscillations. The minimum decay rate is observed at the resonance frequency $\Omega = \gamma_0$, where the antisymmetric mode is predominantly excited and radiative losses are suppressed.

## Conclusions

Thus, in a system of two closely spaced, identical conducting rods, an incident ultrashort terahertz pulse can be efficiently converted into nearly monochromatic radiation. The energy of the resulting monochromatic wave can noticeably exceed the energy carried by the incident pulse within the corresponding spectral region. This spectral redistribution results from the interference between the radiation emitted by the two rods, which act as coupled oscillators. In such a system, the radiation emitted by each oscillator at any given moment depends significantly on the state of its neighboring oscillator. Within the coupled-oscillator framework, this leads to a modified definition of the relaxation coefficient. In addition to the theoretical analysis, the predicted behavior is confirmed by numerical simulations.

**Funding:** This work was supported by the Scientific Research Grant of the Higher Education and Science Committee of the MESCS of Armenia (21AG-1C061).

**Competing interests:** The authors declare no competing interest

# Supplementary Information for

## Relaxation characteristics and spectral compression in the system of coupled radiating oscillators

Henrik Parsamyan,* Hasmik Shahinyan, and Khachatur Nerkararyan

*Institute of Physics, Yerevan State University, A. Manoogian 1, Yerevan, 0025, Armenia*

Corresponding author*: hparsamyan@ysu.am*

*The Supplementary Information contains:* Details of the derivation of the governing differential equations (Sec. S1); Derivation of the relation of relaxation coefficient (Sec. S2); Details of numerical simulations (Sec. S3).

### S1. DERIVATION OF THE GOVERNING DIFFERENTIAL EQUATIONS

The differential equations governing the dipole moments $D_1(t)$ and $D_2(t)$ of two identical, coupled radiating oscillators can be given by [1]:

$$\ddot{D}_1 + 2\gamma\dot{D}_1 + \omega_0^2 D_1 = \alpha\mathcal{E}_0(t)cos\omega t + 2\omega_0\mu D_2 \quad \text{(S1)}$$

$$\ddot{D}_2 + 2\gamma\dot{D}_2 + \omega_0^2 D_2 = 2\omega_0\mu D_1 \quad \text{(S2)}$$

Here, $\omega_0$ denotes the resonant frequency of an isolated oscillator, $\gamma$ is the yet-to-be-determined relaxation parameter, $\mu$ is the coupling coefficient, $\alpha$ characterizes the interaction of the first oscillator with the external field, $\mathcal{E}_0$ is the time-dependent amplitude of the external electric field, and $\omega$ is the carrier frequency of the field.

We seek solutions to the system of equations in the following form:

$$D_1(t) = d_{11}(t)cos\omega t + d_{12}(t)sin\omega t \quad \text{(S3)}$$

$$D_2(t) = d_{21}(t)cos\omega t + d_{22}(t)sin\omega t \quad \text{(S4)}$$

We consider the system under the following conditions:

$$\gamma, |\omega_0 - \omega|, \mu \ll \omega_0 \quad \text{(S5)}$$

Furthermore, it is assumed that the field amplitude $\mathcal{E}_0(t)$ varies slowly in time compared with the period of the carrier frequency. This slowly varying envelope approximation allows us to neglect the second-order derivatives of the $d_{ik}(t)$ amplitudes. Substituting equations (S3) and (S4) into equations (S1) and (S2), and separating the $cos\omega t$ and $sin\omega t$ terms, we obtain:

$$\dot{d}_{11} + \gamma d_{11} + \Omega d_{12} + \mu d_{22} = 0 \quad \text{(S6)}$$

$$\dot{d}_{12} + \gamma d_{12} - \Omega d_{11} - \mu d_{21} = \frac{\alpha}{2\omega_0}\mathcal{E}_0(t) \quad \text{(S7)}$$

$$\dot{d}_{21} + \gamma d_{21} + \Omega d_{22} + \mu d_{12} = 0 \quad \text{(S8)}$$

$$\dot{d}_{22} + \gamma d_{22} - \Omega d_{21} - \mu d_{11} = 0 \quad \text{(S9)}$$

with $\Omega = \omega - \omega_0$:

In accordance with coupled-mode theory, renormalization of the $p_{ik}(t)$ amplitudes transforms equations (S6) - (S9) into equations (4)-(7) in the paper.

To determine the resonant frequencies of the system, we neglect the external field and relaxation processes, assuming that in this case, the $d_{ik}(t)$ amplitudes are time independent. Under these conditions, the system of equations admits two solutions:

1. $\Omega = -\mu$, $d_{11} = d_{21}$ and $d_{12} = d_{22}$, $\omega_s = \omega_0 - \mu$ (S10)

   The oscillators oscillate in phase, and the net dipole moment of the system represents the sum of the individual dipole moments (symmetric mode).

2. $\Omega = \mu$, $d_{11} = -d_{21}$ и $d_{12} = -d_{22}$, $\omega_{as} = \omega_0 + \mu$ (S11)

   The oscillators oscillate out-of-phase, and the net dipole moment of the system is zero (antisymmetric mode).

In the main text of the article, the $d_{ik}$ dipole moments are normalized as.

$$p_{ik} = d_{ik}\sqrt{\frac{\omega_0^4}{6\gamma_0 c^3}} \tag{S12}$$

## S2. DERIVATION OF THE RELAXATION COEFFICIENT

For an isolated oscillator with dipole moment amplitude $p_0$ and oscillation frequency $\omega_0$, the total radiated power $\wp_{rad}$ and stored electromagnetic energy $W$ are given by [2]:

$$\wp_{rad} = \frac{\omega_0^4}{3c^3} d_0^2 \tag{S13}$$

$$W = \frac{1}{2\gamma_0}\frac{\omega_0^4}{3c^3} d_0^2 = \frac{1}{2\gamma_0}\wp_{rad} \tag{S14}$$

with $\gamma_0$ being the relaxation coefficient of an isolated oscillator.

The relationship between the net radiated power and the total energy also follows directly from their physical definitions. The relaxation coefficient is therefore determined from the following relation [1]:

$$\frac{dW}{dt} = -2\gamma_0 W \tag{S15}$$

On the other hand, the rate of change of the energy stored in the oscillator is determined by the radiated power [1]:

$$\frac{dW}{dt} = -\wp_{rad} \tag{S16}$$

Relations (S14) and (S15) are universal and can also be applied to a system of two closely spaced oscillators with slowly varying dipole moment amplitudes $d_1(t)$ and $d_2(t)$.

Neglecting the relatively small coupling energy between the oscillators, we obtain the following expression for the total energy of the system:

$$W_t = W_1 + W_2 = \frac{\omega_0^4}{6\gamma_0 c^3}(d_1^2 + d_2^2) \quad \text{(S17)}$$

However, under the given conditions, the dipolar radiation is determined by the sum of dipole moment amplitudes of the oscillators $p_t = p_1 + p_2$. Consequently, the total power radiated by the system is given by:

$$\wp_{t,rad} = \frac{\omega_0^4}{3c^3}(d_1 + d_2)^2 \quad \text{(S18)}$$

The total relaxation coefficient $\gamma$ of the system can therefore be expressed as:

$$\gamma = \frac{1}{2}\frac{\wp_{t,rad}}{W_t} = \gamma_0 \frac{(d_1+d_2)^2}{d_1^2+d_2^2} \quad \text{(S19)}$$

## S3. DETAILS OF NUMERICAL SIMULATIONS

The rods, each of diameter 10 μm and length 100 μm, are positioned parallel to one another with a surface-to-surface separation of 3 μm along the propagation direction ($k_x$) of the driving pulse. The excitation is a linearly polarized ultrashort pulse ($E_z$) with a center frequency of 1.31 THz and a duration of 2.2 ps, defined as the full width at half maximum (FWHM) of the power temporal spectrum. The offset is 15 ps. The rods were assigned a complex refractive index of 447 + 534i, corresponding to typical reported values of the optical constants of gold in the THz spectral range near 1.3 THz [3]. The system was excited with a scattering-type source. A standard pulse excitation was used in all simulations, with the source characterized by its central frequency and pulse duration. To ensure sufficient spatial resolution near the structures, a rectangular prism-shaped mesh override region with dimensions matching that of the scattering-type source (50 μm × 50 μm × 200 μm) was defined around the rods. Within this region, the mesh step was set to 0.2 μm along the $x$ and $y$ directions and 5 μm along the $z$ direction. The non-uniform mesh was used elsewhere with an accuracy factor of 3. The entire structure was enclosed by perfectly matched layers to suppress reflections of electromagnetic waves from the computational boundaries.